\documentclass[prb,showpacs,floatfix,preprint,amsmath,amssymb]{revtex4}

\usepackage{graphicx}
\usepackage{dcolumn}
\usepackage{bm}
\usepackage{epsfig}

\newcommand{\fig}[1]{Fig.~\ref{#1}}

\begin{document}
\title{Combining Molecular Dynamics with Lattice--Boltzmann: A Hybrid 
Method for the Simulation of (Charged) Colloidal Systems}
\author{Apratim Chatterji}
\author{J\"urgen Horbach}\email{horbach@uni-mainz.de}
\affiliation{Institut f\"ur Physik, 
             Johannes Gutenberg--Universit\"at Mainz,
             D--55099 Mainz, Germany}

\centerline{\underline{J. Chem. Phys. {\bf 122}, 184903 (2005)}}

\begin{abstract}
We present a hybrid method for the simulation of colloidal systems,
that combines molecular dynamics (MD) with the Lattice--Boltzmann (LB)
scheme. The LB method is used as a model for the solvent in order to
take into account the hydrodynamic mass and momentum transport through
the solvent. The colloidal particles are propagated via MD and they are
coupled to the LB fluid by viscous forces. With respect to the LB fluid,
the colloids are represented by uniformly distributed points on a sphere.
Each such point (with a velocity ${\bf V}({\bf r})$ at any off--lattice
position ${\bf r}$) is interacting with the neighboring eight LB nodes by
a frictional force ${\bf F}= \xi_0 ({\bf V}({\bf r})- {\bf u}({\bf r}))$
with $\xi_0$ being a friction force and ${\bf u}({\bf r})$ being the
velocity of the fluid at the position ${\bf r}$. Thermal fluctuations are
introduced in the framework of fluctuating hydrodynamics. This coupling
scheme has been proposed recently for polymer systems by Ahlrichs and
D\"unweg [J. Chem. Phys. {\bf 111}, 8225 (1999)].  We investigate several
properties of a single colloidal particle in a LB fluid, namely the
effective Stokes friction and long time tails in the autocorrelation
functions for the translational and rotational velocity.  Moreover,
a charged colloidal system is considered consisting of a macroion,
counterions and coions that are coupled to a LB fluid.  We study the
behavior of the ions in a constant electric field. In particular, an
estimate of the effective charge of the macroion is yielded from the
number of counterions that move with the macroion in the direction of
the electric field.
\end{abstract}


\maketitle

\section{Introduction}

In colloidal suspensions, big molecules (colloids) with a typical
size of the order of 10\,nm to several $\mu$m are immersed in an
atomistic solvent~\cite{russel,review}. A detailed modeling of such
systems would be provided by a molecular dynamics (MD) simulation
where both colloids and solvent atoms are propagated via Newton's
equations of motion. However, this approach has severe drawbacks due
to the large size difference between colloids and solvent particles.
One would have to take into account the microscopic details of a large
amount of solvent particles which are irrelevant on the typical length
and time scale of the colloids. In order to circumvent this problem,
one may describe the interactions between the colloids by an effective
potential, thus avoiding the explicit consideration of the solvent's
degrees of freedom~\cite{review}.  But then the hydrodynamic mass
and momentum transport through the solvent is completely neglected
and hence, hydrodynamic interactions between colloidal particles are
not taken into account. But it is well--known that many transport
properties in colloidal suspensions are affected by hydrodynamic
interactions~\cite{russel,review}.

In recent years many efforts have been undertaken to model colloidal
suspensions by mesoscopic simulation techniques. The idea is to describe
the solvent on a coarse--grained level, whereby the properties of
the solvent on hydrodynamic time and length scales are correctly
recovered.  Different approaches exist in the literature that range
from particle--based methods such as dissipative particle dynamics
(DPD)~\cite{groot97,espanol99,lowe99,pago01} and stochastic rotation
dynamics (SRD)~\cite{malevanets99,lamura01,ripoll04,kikuchi02,kikuchi03}
to the Lattice--Boltzmann (LB) method~\cite{ladd94_1,ladd94_2,succi02}
where the Navier--Stokes equations are solved on a lattice via a kinetic
equation. All these methods have their advantages and disadvantages,
and they might provide complementary information for a given problem. 

In this paper, we present a hybrid MD/LB method for the simulation of
colloidal systems. In this approach, the coupling between colloids
and solvent (LB fluid) is realized locally at the surface of the
colloids. The modeling of the solvent by a LB fluid has the advantage that
hydrodynamic properties of the solvent (e.g., the shear viscosity $\eta$)
are incorporated as input parameters, and thus they can be varied over
a broad range (compared to particle--based methods). Moreover, the LB
method allows to implement easily the rotational degrees of freedom of
the colloids, an issue that is difficult to handle in particle--based
models of the solvent.

Several methods have been proposed for the introduction
of colloidal particles in a LB fluid.  In Ladd's
method~\cite{ladd94_1,ladd94_2,nguyen02}, a colloid is represented by
its surface which cuts some of the links between lattice nodes of the LB
fluid.  Halfway along these links, boundary points are placed where the
one--particle distribution functions, that represent fluid populations,
are bounced back such that no--slip boundary conditions are obtained. A
different approach for the particle--fluid coupling has been proposed by
Lobaskin {\it et al.}~\cite{lobaskin04_1} Here, a colloid is represented
by point particles that are connected with each other by springs (modeled
by a FENE potential).  The latter point particles interact with the fluid
locally via a friction force~\cite{kalthoff96,ahlrichs99,ahlrichs01}.
The coupling scheme that we present in this paper can be seen as one in
between Ladd's bounce back rules and the frictional coupling scheme of
Lobaskin {\it et al.} As in Ladd's method, each colloid is represented
by boundary points on its surface and the total force and torque that
is exerted on a colloid by the fluid is obtained by summing up all the
contributions from the boundary points.  Different from Ladd's method
we use local friction forces between the boundary points and the fluid
as proposed by Lobaskin {\it et al.} However, no interaction potential
between boundary points is required in our scheme.  We also avoid some
inherent problems of Ladd's method. Our scheme does not introduce shape
fluctuations of a moving colloidal particle as well as mass fluctuations
in the fluid that are induced by the bounce back rules on a moving
colloidal particle.

Many attempts have been devoted very recently to the development
of simulation methods for charged colloids, using the LB method or
finite difference schemes to model hydrodynamic interactions. These
approaches consider charged colloids in the framework of the primitive
model~\cite{review}, i.e.~as a system of negatively charged macroions,
small counterions of positive charge and small coions of negative charge.
In Refs.~\cite{horbach01,capuani04,kodama04,kim04} the small ions are
modeled as charge densities on the level of the Poisson--Boltzmann
equation.  The propagation of these charge densities is achieved by
the so--called electrokinetic equations~\cite{russel} that couple
the Navier--Stokes equations with a dynamic generalization of the
Poisson--Boltzmann equation. The description of charged colloids by the
electrokinetic equations has some drawbacks: Of course, correlations
between the ions are neglected, since the Poisson--Boltzmann equation
is a mean--field description. Moreover, it is not clear how one can
properly introduce thermal fluctuation in the electrokinetic equations
and thus, so far only calculations at zero temperature are possible.
Therefore, Lobaskin {\it et al.}~\cite{lobaskin04_2} have followed a
different strategy by considering explicitly small ions that are moved
together with macroions by their hybrid MD/LB scheme.  Using MD/LB
approach, we also model charged colloidal systems in this way, and,
as an application we consider below the problem of a charged colloidal
system in an external electric field.

The rest of the paper is organized as follows: In the next section,
we give a brief introduction to the LB method. Then, we present our
scheme to couple colloidal particles with a LB fluid (Sec. III) and the
use of this scheme in a hybrid MD/LB method (Sec. IV). Applications are
shown in Sec. V where a neutral colloid in a LB fluid is considered as
well as a macroion and small ions in a LB fluid that are subject to an
external electric field. Finally, we summarize and discuss the method
and the results.

\section{Lattice Boltzmann method}

In this section, we briefly describe the LB method used in
our work.  A detailed description can be found in several review
articles~\cite{ladd94_1,ladd94_2} and in the book by Succi~\cite{succi02}.

In the LB method a discretized version of a simple kinetic equation is
solved numerically on a lattice.  The central quantity in this equation
is an one--particle distribution function $n_i({\bf r},t)$ which gives
the particle population on a lattice node ${\bf r}$ at time $t$ with a
discrete velocity ${\bf c}_i$. The discrete space of velocities $\{{\bf
c}_i\}$ has to be constructed such that the underlying kinetic equation
is consistent with the Navier--Stokes equations. In particular, the
$\{{\bf c}_i\}$ should not introduce any artificial anisotropies in the
hydrodynamic equations. Several choices for $\{{\bf c}_i\}$ are possible.
In this work, the velocity space consists of 18 vectors placed on lattice
nodes of a cubic lattice, 6 of these point to the nearest neighboring nodes
and 12 to the next--nearest neighboring nodes. These velocity vectors have
the absolute values 1 $a/\tau$ and $\sqrt{2}$ $a/\tau$, respectively,
with $a$ being the lattice spacing and $\tau$ the elementary time unit.
Note that the latter velocity space can be constructed by projecting the
24 unit vectors of a four dimensional FCHC lattice onto three dimensions.

The equations of motion for the $n_i({\bf r},t)$ can be written as
\begin{equation}
   n_i({\bf r}+{\bf c}_i \tau, t+\tau) =  n_i({\bf r},t)
    + \Delta_i [\{ n_i({\bf r},t) \} ],
\label{eq1}
\end{equation}
where the collision operator $\Delta_i$ describes the change of $n_i$
due to collisions between the particles (of course $\Delta_i$ has to be
further specified).  The moments of $n_i({\bf r},t)$ are hydrodynamic
fields, namely the mass density
\begin{equation}
  \rho ({\bf r},t) = \sum_{i=1}^{18} n_i ({\bf r},t) \ ,
\label{eq2}
\end{equation}
the momentum density 
\begin{equation}
    {\bf j} ({\bf r},t) = \sum_{i=1}^{18} n_i ({\bf r},t) {\bf c}_i \ ,
\label{eq3}
\end{equation}
and the momentum flux
\begin{equation}
  {\bf \Pi} ({\bf r},t) = \sum_{i=1}^{18} n_i ({\bf r},t) 
   {\bf c}_i {\bf c}_i \ .
\label{eq4}
\end{equation}
Note that ${\bf j}$ can be also written as ${\bf j}\equiv\rho {\bf u}$
with ${\bf u}$ denoting the flow velocity.

The aim is to construct the collision operator $\Delta_i$ in as
simple a form as possible such that the time evolution for the moments in
Eqs.~(\ref{eq1})--(\ref{eq4}) is consistent with the Navier--Stokes
equations. To this end, we assume that $\Delta_i$ describes small
deviations from equilibrium. Hence, it can be given in a linearized
form,
\begin{equation}
   \Delta_i [\{ n_i({\bf r},t) \} ] = \sum_{j=1}^{18} L_{ij}
           (n_j({\bf r},t) - n_j^{\rm eq}) \ ,
\label{eq5}
\end{equation}
where $n_i^{\rm eq}$ is the local equilibrium distribution function
and $L_{ij}$ denotes a matrix element of the collision operator. 

The equilibrium population $n_i^{\rm eq}$ can be expressed
as~\cite{ladd94_1,succi02}
\begin{equation}
   n_i^{\rm eq} = a^{c_i} \left[\rho + 
                   \frac{1}{c_{\rm s}^2} \rho {\bf u} \cdot {\bf c}_i 
                    + \frac{1}{c_{\rm s}^4} \rho {\bf u} {\bf u} 
                   : ({\bf c}_i {\bf c}_i - c_s^2 {\bf 1})
                   \right] \ ,
\label{eq6}
\end{equation}
where $n_i^{\rm eq}$ depends only on $\rho$ and ${\bf u}$. The parameter
$c_{\rm s}$ in Eq.~(\ref{eq6}) denotes the sound velocity. The values
for $c_{\rm s}$ as well as for the coefficients $a^{c_i}$ can be fixed
by a Chapman--Enskog expansion such that the Navier--Stokes equations
in the limit of small ${\bf u}$ are recovered. This yields $c_{\rm
s}=\sqrt{1/2}$. The parameters $a^{c_i}$ can have the two values $a^{1}$
and $a^{\sqrt{2}}$ where the superscript corresponds to the absolute
value of the velocity vector. In our case, we have $a^{1}=1/12$ and
$a^{\sqrt{2}}=1/24$.  The conservation of mass and momentum requires
that $\rho$ and ${\bf j}$ are unaffected by the collision operator. The
kinematic viscosity $\nu$ is related to the eigenvalue $\lambda$ of
$L_{ij}$ that corresponds to the eigenvector $c_{i\alpha}c_{j\beta}$
(herein, the indices $\alpha$, $\beta$ with $\alpha \neq \beta$ denote
the Cartesian components):
\begin{equation}
  \nu = - \frac{1}{6} \left( \frac{2}{\lambda} + 1 \right) \frac{a^2}{\tau}
\label{eq7}
\end{equation}
In the following we choose $\lambda=-1.75$ which corresponds to 
$\nu = 0.0238$ $a^2/\tau$.

All non--hydrodynamic modes (ghost modes), that are due to the use of
a lattice, are suppressed by setting the corresponding eigenvalues of
$L_{ij}$ to -1.  Since we are interested in incompressible LB fluids also
the eigenvalue related to the bulk viscosity is chosen to be -1.

As shown by Ladd~\cite{ladd93}, the LB scheme as defined by
Eqs.~(\ref{eq1})--(\ref{eq7}) allows also the introduction of thermal
fluctuations. This can be done in the framework of fluctuating
hydrodynamics~\cite{landau}. To this end, one adds an additional
stochastic term $n_i^{\prime}$ in Eq.~(\ref{eq1}) which represents the
addition of thermal fluctuations to the stress tensor,
\begin{equation}
  n_i^{\prime} = - \frac{a^{c_i}}{c_{\rm s}^4} \sum_{\alpha \beta}
                   \sigma_{\alpha \beta}^{\prime}
                   c_{i \alpha} c_{i \beta} \ .
  \label{eq8}
\end{equation}
The random stresses $\sigma_{\alpha \beta}^{\prime}$ are random numbers
with zero mean and white noise behavior, i.e.
\begin{eqnarray}
 & & \langle \sigma_{\alpha \beta}^{\prime} ({\bf r},t)
  \sigma_{\gamma \delta}^{\prime} ({\bf r}^{\prime},t^{\prime}) 
   \rangle  \nonumber \\
 & & =  A \delta_{{\bf r} {\bf r}^{\prime}} \delta_{t t^{\prime}}
    \left( \delta_{\alpha \gamma} \delta_{\beta \delta} +
        \delta_{\alpha \delta} \delta_{\beta \gamma} -
      \frac{2}{3} \delta_{\alpha \beta} \delta_{\gamma \delta} \right) 
  \label{eq9}
\end{eqnarray}
where the $\delta$'s are Kronecker deltas and $A$ is a constant which
has to be chosen such that the fluctuation--dissipation theorem holds.
The latter condition requires the choice $A = 2 \eta k_B T \lambda^2$.

An elementary step in a LB simulation can be decomposed in a collision
step and a propagation step.  In the collision step, the interaction
between the ``particles'' at node ${\bf r}$ at the collision time
$t^{\star}$ is taken into account that results in the postcollision
function
\begin{equation}
   n_i^{\star}({\bf r},t^{\star}) = 
   n_i({\bf r},t^{\star})
   + \Delta_i [\{ n_i \} ] + n_i^{\prime}({\bf r},t^{\star}).
 \label{eq10}
\end{equation}
Using the $n_i^{\star}({\bf r},t^{\star})$, the $n_i$ can be updated
in the propagation step,
\begin{equation}
  n_i({\bf r}+{\bf c}_i,t+1) = n_i^{\star}({\bf r},t^{\star}) \ .
  \label{eq11}
\end{equation}
In our implementation we have omitted non--linear terms in ${\bf u}$ and
thus, we describe a fluid on the level of the linearized Navier--Stokes
equations (for more details see Refs.~\cite{ladd94_1,ladd94_2}).

\section{The coupling of the LB fluid to colloidal particles}

Consider a Brownian particle of mass $M$ in a solvent.  In first
approximation, one can describe the motion of the particle without
taking into account explicitly the solvent particles. To this end, one
assumes that, on the typical Brownian time scale, the collisions of the
particle with the solvent particles can be modeled by Gaussian random
forces ${\bf f}_{\rm r}$.  These forces lead to a systematic friction
force $-\xi_0 {\bf V}(t)$ on the particle where $\xi_0$ is the friction
coefficient and ${\bf V}(t)$ the velocity of the particle at time $t$.
The resulting equation of motion is a Langevin equation (see, e.g.,
Ref.~\cite{russel}),
\begin{equation}
   M \frac{d^2 {\bf r}}{dt^2} = 
   {\bf F}_{\rm c} -\xi_0 {\bf V}(t) + {\bf f}_{\rm r} \ .
   \label{langevin}
\end{equation}
Here, ${\bf r}$ is the position of the particle and ${\bf F}_{\rm
c}$ denotes a conservative force due to the interaction with other
particles. The Cartesian components of the forces, $f_{{\rm r},\alpha}$,
are random numbers that are uncorrelated with zero mean, i.e.
\begin{eqnarray}
  \langle f_{{\rm r}, \alpha}({\bf r}, t) \rangle & = & 0 \\
  \langle f_{{\rm r}, \alpha}({\bf r}, t)
          f_{{\rm r}, \beta}({\bf r}^{\prime}, t^{\prime}) 
   \rangle & = & A \delta_{\alpha \beta} \delta ({\bf r} - {\bf r}^{\prime} )
                   \delta (t - t^{\prime}) \; .
\end{eqnarray}
The amplitude $A$ is given by the fluctuation--dissipation theorem, $A =
2 k_B T \xi_0$.

In Eq.~(\ref{langevin}) the interaction of the Brownian particle with the
solvent is described by the force ${\bf F}_{\rm B} = -\xi_0 {\bf V}(t)
+ {\bf f}_{\rm r}$.  By this force, transport of momentum in the fluid
due to the motion of the particle is not taken into account and thus
one does not recover the correct hydrodynamic behavior of the Brownian
particle. This is reflected, e.g., in the behavior of the velocity
autocorrelation function $C_v(t)$. For a single particle, propagated
according to Eq.~(\ref{langevin}) (with ${\bf F}_{\rm c} = 0$), $C_v(t)$
decays exponentially, $C_v(t) \propto \exp(-\xi_0 t/M)$, whereas the
correct hydrodynamic behavior is a power law decay at long times, $C_v(t)
\propto t^{-3/2}$ (the so--called long--time tail)~\cite{russel}. However,
in order to incorporate hydrodynamics, Eq.~(\ref{langevin}) can be
easily modified by coupling the Brownian particle to a LB fluid. The
essential point is to replace the {\it absolute} velocity ${\bf V}(t)$
of the particle in the frictional force term by its velocity relative
to the fluid. The force ${\bf F}_{\rm B}$ is then modified to
\begin{equation}
   {\bf F}_{\rm B} = - \xi_0 
        \left( {\bf V}({\bf r}, t) - {\bf u}({\bf r}, t) \right) 
       + {\bf f}_{\rm r}
   \label{eq_cf}
\end{equation}
where ${\bf u}({\bf r}, t)$ is the velocity of the fluid at the position
of the particle. Since the position of the particle is continuous
in space, we have to use an interpolation scheme to determine ${\bf
u}({\bf r},t)$. As in Ref.~\cite{ahlrichs99,ahlrichs01}, we use a
linear interpolation scheme that estimates ${\bf u}({\bf r},t)$ from
the eight nearest lattice nodes around ${\bf r}$.  Of course, Newton's
third law requires that whenever the Brownian particle is subject to
a force ${\bf F}_{\rm B}$ as given by Eq.~(\ref{eq_cf}), the force
$-{\bf F}_{\rm B}$ has to be applied to the fluid nodes with which the
particle interacts (see Refs.~\cite{ahlrichs99,ahlrichs01}).
One may wonder why fluctuations have to be added both to the LB fluid
and the Brownian particle. However, it was shown in Ref.~\cite{ahlrichs98}
that only then the fluctuation--dissipation theorem holds for the total system
of a particle in a LB fluid coupled via Eq.~(\ref{eq_cf}).

Up to now we have considered the Brownian particle as a point particle
which has, in particular, no rotational degrees of freedom (however,
its effective hydrodynamic radius with respect to the LB fluid is of
the order of the lattice spacing $a$).  In order to model an extended
object, such as a sphere of radius $R$, the force coupling as given
by Eq.~(\ref{eq_cf}) can be generalized as follows: As in Ladd's
method~\cite{ladd94_1,ladd94_2} the sphere is represented by boundary
points on its surface, whereby the surface is permeable for the LB
fluid. In our method, the boundary points are placed uniformly on the
surface of the sphere. The uniform distribution of boundary points can
be achieved by the following iterations: One starts from points that are
placed at the corners of an octahedron.  The envelope of this octahedron
is a sphere of radius $R$, the centre of which is denoted by
${\bf R}_{\rm cm}$ in the following. Then one places new points
halfway along the twelve edges of the octahedron and shifts these points
such that they sit on the surface of the latter sphere.  As a result,
a polygon with 18 corners and 48 edges is obtained.  In this object,
again new points are placed halfway along the edges and shifted to the
surface of the enveloping sphere. This object has now 66 corners, i.e.~66
boundary points, and is used in this work as a model for a colloidal
particle. A snapshot is shown in Fig.~\ref{fig1}. We consider that each
of the boundary nodes has a mass $M/66$ where $M$ is the mass
of the colloidal particle. Each of the boundary
points is coupled to the LB fluid according to Eq.~(\ref{eq_cf}). The
total force on the particle is then determined by the sum
\begin{equation}
  {\bf F}_{\rm B, tot} = \sum_{i_{\rm b}=1}^{66} 
  {\bf F}_{\rm B}({\bf r}_{i_{\rm b}}) 
\end{equation}
where ${\bf r}_{i_{\rm b}}$ denotes the positions of the 66 boundary
points. The torque that is exerted on the particle by the fluid is
updated via
\begin{equation}
  {\bf T}_{\rm B} = \sum_{i_{\rm b}=1}^{66} 
  {\bf F}_{\rm B}({\bf r}_{i_{\rm b}}) \times 
  ({\bf r}_{i_{\rm b}} - {\bf R}_{\rm cm}) \; .
\end{equation}
${\bf F}_{\rm B, tot}$ and ${\bf T}_{\rm B}$ can be used to update the
translational and the rotational velocity of the colloidal particle which
we denote by ${\bf V}_0$ and ${\bf \omega}_0$, respectively. The simplest
discretized form of the equations of motion is the Euler algorithm
which is of first order in the time step $h$ for the integration,
\begin{eqnarray}
  {\bf V}_0 (t+ h) & = & {\bf V}_0 (t) + {\bf F}_{\rm B, tot} 
  \frac{h}{M} \; \label{euler1} \\
  {\bf \omega} (t+ h) & = & {\bf \omega} (t) + {\bf T}_{\rm B} 
  \frac{h}{I} \; .
  \label{euler2}
\end{eqnarray}
In Eq.~(\ref{euler2}), $I$ is the moment of inertia of the particle.
The Euler algorithm has of course very bad properties with respect to
stability, temperature drift etc. In the next section, we present an
integrator based on the Heun algorithm~\cite{heun1900} that is of second
order in $h$.

The boundary points that we use for the representation of the surface of
the spherical particle are not rotated according to the updated rotational
velocity $\omega_0$. These points are fixed with respect to the centre
of mass position ${\bf R}_{\rm cm}$ of the particle. The idea is that
the surface of a spherical particle can be always represented by the
same set of uniformly distributed points if the surface grid that they form
is fine enough. We have checked that the sphere with 66 boundary points
(see Fig.~\ref{fig1}) is sufficient for particle radii up to $R=5.0 a$.
In this case, no change in any physical properties is seen if the
particle is decorated with more boundary points.

\section{The hybrid MD/LB scheme}

Now, we discuss the algorithm that we use for the integration of
the equations of motion of a system of colloidal particles coupled
to a LB fluid.  To this end, we use a generalized velocity Verlet
algorithm~\cite{frenkel96}.  For the colloidal particles we use the model
with 66 boundary points that we have described in the previous section.

The force on particle $i$ can be decomposed into three terms,
\begin{equation}
  {\bf F}_{{\rm tot},i}=
  {\bf F}_{{\rm c},i} 
  + {\bf F}_{{\rm f},i} + {\bf F}_{{\rm r},i},
\end{equation}
where ${\bf F}_{{\rm c},i}$ is the contribution from conservative forces.
${\bf F}_{{\rm f},i}$ and ${\bf F}_{{\rm r},i}$ denote friction and
random forces, respectively.  To compute ${\bf F}_{{\rm f},i}$, one sums
up the contributions from the 66 boundary points,
\begin{equation}
  {\bf F}_{{\rm f},i} = - \sum_{i_{\rm b} = 1}^{66}
  \xi_{0,i_{\rm b}} \left[ {\bf v}_i 
              + \mathbf{\omega} \times 
                   \left( {\bf r}_{i_{\rm b}} - {\bf r}_i\right)
              - {\bf u}({\bf r}_{i_{\rm b}}) \right] \, ,
\end{equation}
where ${\bf r}_i$ and ${\bf v}_i$ are respectively the position and the
velocity of the centre of mass of particle $i$, ${\bf r}_{i_{\rm b}}$
is the position of the boundary node $i_{\rm b}$ of particle $i$, ${\bf
\omega}$ is its angular velocity, and the fluid velocities at the boundary
points are denoted by ${\bf u}({\bf r}_{i_{\rm b}})$.  $\xi_{0,i_{\rm b}}$
is the friction coefficient of boundary point $i_{\rm b}$ and thus the
total friction coefficient $\xi_0$ is given by $\xi_0 = \sum_{i_{\rm
b}} \xi_{0,i_{\rm b}}$. In the following, each of the 66 $\xi_{0,i_{\rm b}}$
is assigned the same value $\xi_0/66$ for a given particle.

The random force can be written as
\begin{equation}
   {\bf F}_{{\rm r},i} = \sqrt{2 \pi k_B T} 
     \sum_{i_{\rm b} = 1}^{66} \sqrt{\xi_{0,i_{\rm b}}} \;
      {\bf \theta}_{i_{\rm b}} \, .
    \label{ran_def}
\end{equation}
Here, the ${\bf \theta_{i}}$ is a vector of random numbers with the
Cartesian components $\theta_{i,\alpha}$ ($\alpha \in \{x, y, z\}$)
for which  
\begin{eqnarray}
  \langle \theta_{i, \alpha}(t) \rangle & = & 0 \label{ran_av} \\
  \langle \theta_{i, \alpha}(t) \theta_{j, \beta}(t^{\prime}) \rangle 
      & = & \delta_{ij} \delta_{\alpha \beta} \delta(t - t^{\prime}) 
    \label{ran_fluc}
\end{eqnarray}
holds. Note that it is not necessary to use random numbers $\theta_{i,
\alpha}$ with a Gaussian distribution for the numerical integration. It
was shown in Ref.~\cite{duenweg91} that it is sufficient to use
uniform random numbers that fulfill the requirements as given by
Eqs.~(\ref{ran_av}) and (\ref{ran_fluc}).

The update of the centre of mass positions of the particles is similar to
that in the velocity Verlet algorithm~\cite{frenkel96}. In an integration
time step $h$ the position change from ${\bf r}_i(0)$ to
\begin{eqnarray}
   {\bf r}_i(h) & = & {\bf r}_i(0) + h {\bf v}_i(0) \nonumber \\
   & & + \frac{h^2}{2M} \left(
        {\bf F}_{{\rm c},i}(0) + {\bf F}_{{\rm f},i}(0) \right)
       + {\bf F}_{{\rm r},i}(h) \, .
  \label{pos_step}
\end{eqnarray}
For the determination of the velocities ${\bf v}_i(h)$, the friction force
${\bf F}_{{\rm f},i}$ at time $t=h$ is needed if we want to apply the
velocity Verlet scheme. But in this case, the problem arises that ${\bf
F}_{{\rm f},i}(h)$ depends itself on ${\bf v}_i(h)$.  This problem
can be solved if we first approximate ${\bf v}_i(h)$ in an Euler step
to obtain
\begin{equation}
   {\bf v}_i^*(h) = {\bf v}_i(0) + \frac{h}{M} 
   \left( {\bf F}_{{\rm c},i}(0)+{\bf F}_{{\rm f},i}(0) \right)
    + \frac{h}{M} {\bf F}_{{\rm r},i}(h) \, ,
   \label{veuler}
\end{equation}
where the star indicates that we use this velocity only for an estimate
of ${\bf F}_{{\rm f},i}(h)$. Also the angular velocity ${\bf \omega}_i
(h)$ is updated by an Euler step,
\begin{equation}
   {\bf \omega}_i (h) = {\bf \omega}_i (0) + 
     \frac{h}{I} \sum_{i_{\rm b}= 1}^{66} {\bf F}_{\rm f} (0)
            \times \left[ {\bf r}_{i_{\rm b}} - {\bf r}_i(h) \right] \, .
   \label{omeuler}
\end{equation}
With Eqs.~(\ref{veuler}) and (\ref{omeuler}) we yield the friction
force at time $t=h$ as
\begin{eqnarray}
  {\bf F}_{{\rm f},i}(h) & = & - \sum_{i_{\rm b} = 1}^{66}
  \xi_{0,i_{\rm b}} [ {\bf v}_i^*(h)
              + {\bf \omega}(h) \times
                   \left( {\bf r}_{i_{\rm b}} - {\bf r}_i(h)\right) 
                \nonumber \\
        & &   - {\bf u}({\bf r}_{i_{\rm b}}) ] \, .
  \label{ffric}
\end{eqnarray}
With Eq.~(\ref{ffric}) the velocities ${\bf v}_i(h)$ are obtained by
\begin{eqnarray}
   {\bf v}_i(h) & = & {\bf v}_i(0) + \frac{h}{2m}
   [ {\bf F}_{{\rm c},i}(0) + {\bf F}_{{\rm c},i}(h) \nonumber \\
        & &  + {\bf F}_{{\rm f},i}(0) + {\bf F}_{{\rm f},i}(h)
    + 2 {\bf F}_{{\rm r},i}(h) ]
  \label{vel_step}
\end{eqnarray}
Of course, at the same time one has to transfer also the force $-{\bf
F}_{{\rm f},i}(h)-{\bf F}_{{\rm r},i}(h)$ to the appropriate nodes
in the LB fluid.

Our scheme is equivalent to the Heun algorithm~\cite{heun1900}
which is a standard method for the solution of Langevin
equations~\cite{greiner88,paul95}. If we set $\xi_{0,i_{\rm b}}=0$
in Eqs.~(\ref{pos_step}) and (\ref{vel_step}) the algorithm
reduces to the velocity Verlet algorithm for the microcanonical
ensemble~\cite{frenkel96}.

\section{Results}

In this section we present several applications of our MD/LB method.
First, we consider a neutral colloidal particle in a LB fluid to
calculate its effective friction coefficient (part A) and to study long
time tails in the translational and rotational velocity autocorrelation
function (part B). In part C we consider a charged colloid in an electric
field. In the following, we choose $\rho =1.0$~$m_0/a^3$ for the density
and $\eta\equiv \nu \rho=0.02381 a^2 \rho/\tau$ for the shear viscosity
of the LB fluid.

\subsection{Friction coefficient}

Consider a sphere with (hydrodynamic) radius $R_{\rm h}$ that moves through a
viscous fluid due to a gravitational field ${\bf g}$. In the steady state,
it experiences a drag force ${\bf F}_{\rm d}$ which is proportional to
its velocity ${\bf U}$, according to Stokes law~\cite{batchelor}:
\begin{equation}
   {\bf F}_{\rm d} = \xi {\bf U}
\label{stokes_eqn}
\end{equation}
where $\xi$ is the Stokes friction coefficient.  In the case of no--slip
boundary conditions, the friction coefficient is given by $\xi=6 \pi \eta
R_{\rm h}$ (with $\eta$ being the shear viscosity of the fluid). In the steady
state, the force ${\bf F}_{\rm d}$ is equal to the gravitational force
on the sphere,
\begin{equation}
  {\bf F}_{\rm g}= \frac{4}{3} \pi R_{\rm h}^3 (\rho_{\rm p} - \rho) {\bf g},
\end{equation} 
where $\rho_{\rm p}$ and $\rho$ denote the density of the sphere and the
fluid, respectively. Thus, from ${\bf F}_{\rm g}={\bf F}_{\rm d}$ the
friction coefficient $\xi$ can be determined by measuring the velocity
${\bf U}$ of the particle in the steady state.

As a convenient way to determine $\xi$ in a LB simulation, we follow here
the scheme proposed by Ladd~\cite{ladd94_1}. We consider a particle in
a LB fluid. This system is put in a cubic box of volume $V=L^3$ with
periodic boundary conditions in all three Cartesian directions $x$,
$y$, $z$. The particle is held fixed by assigning an infinite mass
to it.  Then, a pressure gradient $\nabla_x p$ is introduced in the
$x$ direction by applying a constant increment $\Delta j_x$ to the $x$
component of the momentum density at each lattice node.  In the steady
state, the total force on the particle is balanced by the sum of the
drag force $F_{{\rm d},x}=V \Delta j_x/ \tau$ and the buoyancy force
$F_{\rm b}=-\frac{4}{3} \pi R^3 \Delta j_x/ \tau$ (remember that $\tau$
denotes the time step of the LB simulation).  Thus the Stokes friction
coefficient is given by $\xi_L = F_{{\rm d},x}/U_x$ where the index $L$
reflects that the particle is moving in a finite system of size $L$. Due
to the periodic boundary conditions, $\xi_L$ describes the friction of
an array of spheres that sit on a cubic lattice with lattice constant
$L$. An analytic expression for $1/\xi_L$ in terms of an expansion of
powers of $1/L$ was first derived by Hasimoto~\cite{hasimoto59} assuming
no--slip boundary conditions. It has the following form:
\begin{equation}
\frac{1}{\xi_L} = \frac{1}{6 \pi \eta} \left(
\frac{1}{R_{\rm h}} - \frac{2.837}{L} 
+ \frac{4.19}{L^3} {R_h}^2 + ... \right)
\label{eq_hasi}
\end{equation}
For our fluid--particle coupling scheme, we can only use this formula
for high values of $\xi_0$. Then, no--slip boundary conditions
are approximately recovered as we shall see in the following.

In Fig.~\ref{fig2}, $1/\xi_L$ is plotted as a function of $1/L$ for
different values of $\xi_0$ from $\xi_0=3.3 \; m_0/\tau$ to $\xi_0=19.8
\; m_0/\tau$.  The radius of the particle is $R=2.5a$ in this case. For
$1/L < 0.04a^{-1}$ the data can be well described by a linear $1/L$
dependence of the form
\begin{equation}
\frac{1}{\xi_L}=\frac{1}{\xi_{\infty}}-B \frac{1}{L} \ ,
\label{eq_fitxi}
\end{equation}
where $\xi_{\infty}$ denotes the friction coefficient for an unbounded
system. Fits with Eq.~(\ref{eq_fitxi}) are shown in Fig.~\ref{fig2}
as dashed lines. In these fits, the slope $B$ changes only slightly
with $\xi_0$: We find $B=6.419 a \tau/m_0$ at $\xi_0=3.3 \; m_0/\tau$ and
$B=6.438 a \tau/ m_0$ at $\xi_0=19.8 \; m_0/\tau$. Moreover, for 
$\xi_0=19.8 \; m_0/\tau$ the data over the full $1/L$ range can be well described by
a fit with Eq.~(\ref{eq_hasi}) (bold solid line in Fig.~\ref{fig2}). In
the latter fit only the hydrodynamic radius $R_{\rm h}$ was used as a fit
parameter for which we find $R_{\rm h}=3.05a$. The good quality of the fit
indicates that at $\xi_0=19.8 \; m_0/\tau$, no--slip boundary conditions are
almost recovered whereas at smaller values of $\xi_0$ mixed stick--slip
boundary conditions are obtained. This can be also inferred from the inset
of Fig.~\ref{fig2} where we have plotted $R_{\rm h}/R$ as determined from
the fitted $\xi_{\infty}$ via $R_{\rm h}/R=\xi_{\infty}/(6 \pi \eta R)$.

One may wonder why the hydrodynamic radius is about 20\% higher than the
assigned radius $R=2.5a$ of the particle. But this is just an artifact
of the discrete nature of the LB fluid. This artifact can be reduced
by increasing the size of the particle. 

The discrete nature of the LB fluid is also reflected in the dependence of
$\xi$ on the size of the particle $R$. In Fig.~\ref{fig3}, we show $\xi$
as a function of $R$ for the two system sizes $L=60a$ and $L=80a$. We see
from the figure that, at small values of $R$, $\xi_L$ does not increase
linearly with the sphere radius $R$ and thus the ratio $R_{\rm h}/R$ is
not a constant in this regime. Only, for $R>3.0 a$, $\xi_L \propto R$
seems to hold to a good approximation.

We have seen that, with the frictional coupling scheme used in this work,
no--slip boundary conditions at the surface of a colloidal particle
can be nearly realized.  For the system considered in this section,
one has to choose a value of $\xi_0$ around $20 \; m_0/\tau$ to obtain
no--slip boundary conditions. Note that the limit of small values of
$\xi_0$ is also of interest. As we shall demonstrate in a forthcoming
publication~\cite{chatt}, the frictional coupling scheme can be also
used to model walls at which mixed stick--slip boundary conditions hold.
This is an important issue for the modeling of fluids in nanoscopic slits.

\subsection{Long time tails}

In this section, we consider again a neutral, spherical particle in
a LB fluid.  Our aim is to study the normalized translational
and rotational velocity autocorrelation function which we denote by
$C_v(t)$ and $C_{\omega}(t)$, respectively.  These functions can be
simply calculated by making use of linear response theory. To this end,
we consider a colloidal particle in a LB fluid at rest. At $t=0$,
we give the particle a translational or rotational velocity. The decay
of these velocities with time, normalized by the initial values, yields
then the functions $C_v(t)$ and $C_{\omega}(t)$, respectively.

Fig.~\ref{fig4} illustrates the short time decay of $C_v(t)$ for
different values of $\xi_0$. The initial decay of these functions is
given by the exponential functions $f(t)=\exp\left( - \frac{\xi_0}{M}
t \right)$ which are shown as dotted lines in Fig.~\ref{fig4} (note that
we have assigned the mass $M=120m_0$ to the particle). Thus, the short
time behavior of $C_v(t)$ is completely controlled by the relaxation
time $\tau_{\rm B}=M/\xi_0$.

As we can also see in Fig.~\ref{fig4}, for $t>\tau_{\rm B}$ the the
average fluid velocity at the surface of the particle essentially equals
the velocity of the particle.  The dashed lines in Fig.~\ref{fig4} are
averaged velocities $v_{\rm s}$ of the fluid at the boundary points of
the particle (normalized by the initial velocity $V(0)$ of the particle).
Obviously, $v_{\rm s}/V(0)$ matches the function $C_v(t)$ at a time
which is of the order of $\tau_{\rm B}$.

For long times one expects the occurrence of a long time tail in $C_v(t)$,
i.e., a decay with the power law $f(t)= A t^{-3/2}$~\cite{alder70}. The
theoretical prediction for the prefactor is $A=M/(12 \rho) [\pi
\nu]^{-3/2}$~\cite{ernst70,cichocki95} (with $\nu=\eta/\rho$ being the kinematic
viscosity of the fluid). According to this prediction the prefactor $A$
does not depend on the details of the particle such as its radius $R$.
The physical origin of the long time tail is the conservation of momentum,
which is transported away diffusively from the particle. Since the
momentum transport in a fluid is spatially long--ranged, one expects the
presence of finite size effects if the decay of velocity correlations
of a colloidal particle in a finite simulation box is considered.

In Fig.~\ref{fig5}, $C_v(t)$ is plotted for different system sizes $L$ for
$\xi_0=6.6 \; m_0/\tau$ and, as a dotted line, the theoretically predicted
long time tail is shown (note that no fit parameters are involved).
We see that the theoretical result match perfectly with the simulation
data. One can also infer from the figure the expected finite size effects.
At a given $L$, $C_v$ seems to approach a constant at long enough times.
This is just a consequence of momentum conservation: At long times the
particle has completely transfered its initial velocity $v_0$ to the LB
fluid and then the whole system moves with a constant velocity $v_0/N_{\rm
tot}$ where $N_{\rm tot}$ is the total number of lattice nodes.

Also the angular velocity correlation function $C_{\omega}(t)$ exhibits
a long time tail but now the exponent for the power law decay is
$-5/2$, i.e.~$f(t)=B t^{-5/2}$. The theoretical prediction for the
prefactor is $B=\pi I/\rho \; [4 \pi \nu]^{-5/2}$~\cite{cichocki00}.
In order to compare the latter theoretical prediction to the simulation
result, one has to estimate the moment of inertia $I$ of the rotating
sphere. On the one hand, $I$ has a contribution $I_1$ from the shell of
the particle that consists of 66 boundary points.  On the other hand,
at long times the moment of inertia is also affected by the rotating
fluid inside the sphere which leads to a second contribution $I_2$
to $I$.  $I_1$ can be estimated by $I_1=\frac{2}{3}MR^2 = 500 \; m_0 a^2$
(i.e., the value of a hollow sphere with radius $R$). We have computed
$I_2$ numerically by considering all the lattice nodes of mass $m_0$
inside the sphere to obtain the value $I_2=228 \; m_0 a^2$. Thus, the
total moment of inertia of the rotating sphere is $I=I_1+I_2=728 \;
m_0 a^2$.  The theoretical prediction using this value of $I$ is shown
in Fig.~\ref{fig6} in comparison with the simulation data for different
values of $L$. Obviously, the theoretical prediction is in nice agreement 
with the numerical data.

As we have already mentioned, we use linear response theory to determine
$C_v(t)$ and $C_{\omega}(t)$ by considering a kicked particle in a LB
fluid which is initially at rest. Of course, we can also calculate
the velocity correlation functions from thermal fluctuations of the
translational and rotational velocity, respectively, and this should
yield identical results. Indeed, this is the case as we can infer from
Fig.~\ref{fig7} and Fig.~\ref{fig8} where $C_v(t)$ and $C_{\omega}(t)$,
respectively, are shown for $L=40a$ and $\xi_0=6.6 \; m_0/\tau$.

We have seen in this section that we recover the the
theoretical predictions for the long time tails in $C_v(t)$ and
$C_{\omega}(t)$. This is in agreement with the study of Lobaskin {\it
et al.}~\cite{lobaskin04_1} who use a slightly different particle--fluid
coupling scheme (see above).

\subsection{Charged Colloids}

Now we demonstrate that our hybrid MD/LB scheme can be applied to charged
colloidal systems. To this end, we consider a charged spherical particle
(macroion) that is immersed in a fluid of small ions and a neutral
``hydrodynamic background'' which is modeled by a LB fluid. This system
is then studied in an electric field to determine the drift velocity
and the effective dynamic charge of the macroion. In the following, we
first introduce the model and the simulation details, before we present
the results of the simulation.

\subsubsection{Model and simulation details}

A potential that describes surprisingly well the effective interactions
between macroions in a charged colloidal suspension is the Debye--H\"uckel
(DH) potential, which is the solution of the linearized Poisson--Boltzmann
equation~\cite{russel}. This potential has a Yukawa form,
\begin{equation}
  u(r)= K \frac{\exp(- r/\lambda_{\rm D})}{r} \ ,
  \label{eq_dh}
\end{equation}
where $\lambda_{\rm D}$ is the so--called Debye screening length and $K$
is a constant depending in particular on the charge of the macroions
$Z_{\rm m}$. Eq.~({\ref{eq_dh}) describes the screening of the Coulomb
interaction between positively charged macroions due to the presence of small ions
in the system, namely negatively charged counterions and positively charged
coions. The screening length is explicitly given by $\lambda_{\rm D}=(4
\pi l_{\rm B} \sum_s z_s^2 \bar{\rho}_s)^{1/2}$ where $l_{\rm B}$ is the
Bjerrum length and $\bar{\rho}_s$ is the average density of microscopic
ions of type $s$ (either counterions or coions).  In principal the DH
potential should only be valid for small charge and surface potential
of the macroions. However, it turns out that also systems with highly
charged particles can be rather well described by the DH
potential if one introduces an effective charge $Z^*_{\rm m}$ for the
macroions. One possibility to determine $Z^*_{\rm m}$ experimentally
is via electrophoresis where the $Z^*_{\rm m}$ is extracted from the
measurement of the electrophoretic mobility (see below). We shall address
here the problem of estimating $Z^*_{\rm m}$ via electrophoresis by
means of our LB/MD simulation method.

To this end, we investigate a charged colloidal system in the framework
of the primitive model~\cite{russel}. Thus, we consider a system of a
positively charged macroion of charge $Z_{\rm m}$ and small ions of charge
$Z_{\rm i}=-1$ (counterions) and of charge $Z_{\rm c}=1$ (coions). Of
course, the number of counterions and coions is chosen such that charge
neutrality of the system holds. The interaction potential between a particle of type
$\alpha$ and a particle of type $\beta$ ($\alpha, \beta = {\rm m, i,
c}$) separated by a distance $r$ from each other is given by
\begin{equation}
  u_{\alpha \beta} = 
  \frac{Z_{\alpha} Z_{\beta} e^2}{4 \pi \epsilon_{\rm r} \epsilon_0 r}
  + A_{\alpha \beta} 
    \exp \left\{ - B_{\alpha \beta} (r - \sigma_{\alpha \beta} ) \right\}
  \label{eq_pot}
\end{equation}
where $e$ is the elementary charge and $\epsilon_{\rm r}$ and $\epsilon_0$
are the reduced dielectric constant (which we set to $\epsilon_{\rm
r}=80$ for water at room temperature) and the vacuum dielectric constant,
respectively. The parameter $\sigma_{\alpha \beta}$ is the distance between two ions
at contact, $\sigma_{\alpha \beta}= R_{\alpha} + R_{\beta}$, where
$R_{\alpha}$ is the radius of an ion of type $\alpha$. In the following, we
use $R_{\rm m}=20$~\AA~and $R_{\rm i}=R_{\rm c}=1$~\AA.  The exponential
in Eq.~(\ref{eq_pot}) is an approximation to a hard sphere interaction
for two ions at contact. For the parameters $A_{\alpha \beta}$ we choose
$A_{\rm mm}=1.84$~eV, $A_{\rm mi}=A_{\rm mc}=0.0556544$~eV, and $A_{\rm
ii}=A_{\rm ic}=A_{\rm cc}=0.0051$~eV. The parameters $B_{\alpha \beta}$
are all set to $4.0$~\AA$^{-1}$. 

We have done simulations for two different systems: The first system is
a mixture of a macroion of charge $Z_{\rm m}=121$ with $471$ counterions
and 350 coions. The second system contains a macroion of charge $Z_{\rm
m}=255$, 555 counterions, and 300 coions. In both cases, the ions
are placed in a cubic simulation box of linear size $L=160$~\AA~using
periodic boundary conditions. All the simulations are done at $T=297$~K.
The Debye screening length $\lambda_{\rm D}$ is for both systems around $7.5$~\AA.
For the masses of the macroion and the small ions we have chosen 60 atomic
units and 4 atomic units, respectively. Thus, the mass of the macroion
is a factor 15 times the mass of a small ion.  The Coulomb part of the
potential and the forces was evaluated by means of Ewald sums with a
constant $\alpha = 0.05$ and by using for the Fourier part of the Ewald
sum all $k$ vectors of magnitude less than 
$k_{\rm c}=2 \pi \sqrt{66}/L$~\cite{frenkel96,allen}.

For the LB fluid to which the ions are coupled we use a cubic lattice
with $40^3$ lattice nodes. Hence, since the size of the simulation box
is $L=160$~\AA, the lattice constant of the LB fluid is $a=4.0$~\AA.
The counterions and coions are treated as point particles with respect to
the LB fluid, i.e., each counterion and coion are equivalent to a sinlge
boundary point on the surface of the colloid. The force on each of the
ions is calculated by Eq.~(\ref{eq_cf}).  The value of the applied random
force on the ion is calculated according to Eq.~(\ref{ran_def}) and is
equal to the random force acting on a single boundary node.  The macroion
is seen by the LB fluid as a sphere of radius $R_{\rm m, LB}=2.5 a$ using
the model with 66 boundary points and $\xi_0=6.6 m_0/\tau$.  The choice
$R_{\rm m, LB}<R_{\rm m}$ is important if systems with more than one
macroion are simulated. In this case the smaller $R_{\rm m, LB}$ prevents
that two macroions get in close contact with respect to the LB fluid.

The equations of motion were integrated with a time step of 1~fs. This
very small time step is necessary because we consider explicitly
counterions and coions as {\it microscopic} particles. A larger
time step could be used if the exponential term in the potential,
Eq.~(\ref{eq_pot}), is replaced by a softer repulsive potential. Moreover,
for the update of the LB fields one could use a larger time step as for
the MD part as it was done in Refs.~\cite{lobaskin04_1,lobaskin04_2}.
However, such optimizations were not necessary for the problems that
are considered in the next section.

\subsubsection{A colloidal particle in an electric field}

The two systems with $Z_{\rm m}=121$ and $Z_{\rm m}=255$ are now studied
in a constant electric field. This leads to a drift velocity $v_{\rm
d}$ of the macroion in the direction of the field (in the following we
apply a field $E_x$ in $x$ direction).  In the linear response regime,
$v_{\rm d}$ is linearly related to $E_x$, $v_{\rm d}= \mu E_x$, where
$\mu$ is the so--called electrophoretic mobility~\cite{russel}. The
latter quantity is of particular interest because, the experimental
determination of $\mu$ allows to extract the effective charge of the
macroion~\cite{hessinger00,wette02,shapran04}.  In the following, we
determine $v_{\rm d}$ as a function of $E_x$ and we estimate the effective
charge of the macroion $Z_{\rm m}^*$ from the number of counterions that
move with the macroion when the electric field is switched on.

We first equilibrated our system for 25000 time steps without the electric
field and without coupling the system to the LB fluid. Fig.~\ref{fig9}
is an equilibrium snapshot of the system with macroion charge $Z_{\rm
m}=255$. One can clearly identify a spherical region around the macroion
(big  sphere), the so--called Debye layer, that contains an excess of
counterions (small light gray spheres) whereas the coions (small black
spheres) are almost excluded from this region.  Far away from the Debye
layer the small ions are randomly distributed.  The presence of the
Debye layer can be also inferred from the radial density distributions
$\rho_{\rm c}$ of counterions and coions around the macroion which are
shown in Fig.~\ref{fig10} for the two systems with different macroion
charges. In the inset of Fig.~\ref{fig10} the instantaneous temperature of the
small ions is shown during a simulation of 10000 time step. We see
that it fluctuates correctly around the assigned value for the temperature,
$T=297$~K.

After the equilibration, the system was coupled to the LB fluid and the
electric field $E_x$ was switched on. We did runs for several values of
$E_x$ ranging from $E_x=0.0025$~V/\AA~to $E_x=0.02$~V/\AA. For each
value of the electric field, runs over $300000$ to $500000$ steps were
done. After reaching the steady state within about 10000 time steps,
positions and velocities of the ions were stored every 500 steps to
obtain the averaged quantities that are presented in the following.

Due to the electric field $E_x$ the macroion and the coions move in the
positive $x$ direction whereas the counterions experience a force in the
opposite direction.  This leads to a distortion of the Debye layer which
gives rise to a force opposing the motion of the macroion. This retarding
force coupled with the viscous drag due to the fluid balances the force
on the macroion due to $E_x$ in the steady state.  Fig.~\ref{fig11}
shows steady state configurations of the ions for the system with
macroion charge $Z_{\rm m}=255$ for two different choices of $E_x$,
namely for $E_x=0.01$~V/\AA~(top) and $E_x=0.02$~V/\AA~(bottom), where
the positive $x$ direction is marked by a black arrow.  It can be seen
from the picture that there are more counterions behind the macroion than
in front of it leading to the distorted counterion charge distribution.
A high value of the electric field (Fig.~\ref{fig11} bottom) strips off
the counterions from the shell moving along with the macroion and hence
the counterion cloud becomes more diffuse.

In order to obtain an estimate of the effective charge of the macroion,
we determine the average number of counterions $N_{\rm c}^*$ that move
along with the macroion, i.e.~those counterions are counted that have a
positive drift velocity in $x$ direction. In the limit of small electric
field $E_x$, i.e.~in the linear response regime, $|Z_i| N_{\rm c}^*$
(here $Z_i=-1$) is a measure of the effective charge $Z_{\rm m}^*$ of the
macroion. Hence, the ratio $N_{\rm c}^*/Z_{\rm m}$ approaches one towards
vanishing values of $E_x$ if $Z_{\rm m}^*$ is equal to $Z_{\rm m}$.
Fig.~\ref{fig12} shows $N_{\rm c}^*/Z_{\rm m}$ as a function of $E_x$
for the two different systems with $Z_{\rm m}=121$ and $Z_{\rm m}=255$.
In both cases, the linear response regime is obviously not approached,
even for small considered values of $E_x$. However, an extrapolation of
the two curves in Fig.~\ref{fig12} to zero electric fields yields values
close to one and thus, we can conclude from the data that $Z_{\rm m}^*$ is
close to the bare charge $Z_{\rm m}$ for the systems under consideration.
We also see in Fig.~\ref{fig12} that the slope for small $E_x$ is smaller
for the system with $Z_{\rm m}=255$ than for the one with $Z_{\rm m}=121$.
This finding shows that, the smaller $Z_{\rm m}$ is, the smaller electric
fields are required to obtain a reasonable estimate of the effective
charge $Z_{\rm m}^*$. That the linear response regime is not reached for
our smallest considered values of $E_x$, is illustrated in the inset of
Fig.~\ref{fig12} which shows the drift velocity $v_{\rm d}$ as a function
of $E_x$. The reason that we did not do simulations for lower values
of $E_x$ is that the drift velocity approaches the order of the thermal
velocity then and thus, it is difficult to yield a reasonable statistics.

The finding $Z_{\rm m}^*\approx Z_{\rm m}$ is in agreement with recent
MD simulations~\cite{shapran04,kreer05,alla98} of similar systems where
the effective charge $Z_{\rm m}^*$ was estimated from the potential of
mean force between the macroions.

\vspace*{0.8cm}
\section{Conclusions}

We have presented a hybrid MD/LB method for the simulation of
colloidal systems that has been applied to several simple colloidal
systems.  Our method can be seen as an alternative to Ladd's coupling
scheme~\cite{ladd94_1,ladd94_2} as well as to the one proposed by Lobaskin
{\it et al.}~\cite{lobaskin04_1,lobaskin04_2} We think that all these
methods (including ours) have their advantages and disadvantages and it
depends on the problem whether one might prefer one or the other method.

We have applied our method to the simulation of charged colloidal
systems where, apart from macroions, counterions and coions are considered
in the framework of the primitive model. For such systems,
our simulation technique has been used to get insight into the properties
of colloids in an external electric field. This is of particular
interest for the understanding of experimental studies where one extracts
effective charges of macroions from the measured electrophoretic
mobility~\cite{hessinger00,wette02,shapran04}. A more extensive study 
on this issue is in preparation.

The model with the explicit consideration of small ions has also a
drawback which was already pointed out in Ref.~\cite{lobaskin04_2}:
Only relatively small size ratios between macroions and small ions are
accessible. Furthermore, only relatively small systems can be simulated
due to the long--ranged character of the electrostatic interactions.
Thus, one looses partly the advantage of the LB method that mesoscopic
length and time scales can be covered. However, as we have mentioned
above, the DH potential describes the effective interactions between
macroions surprisingly well. Therefore, we plan to study systems of,
say, 1000 macroions which interact with each other via an effective
DH potential. This allows then the investigation how the long--time
diffusion of systems of charged colloidal particles at intermediate
densities is affected by hydrodynamic interactions.

{\bf Acknowledgments:}
We thank Burkhard D\"unweg, Vladimir Lobaskin, and Thomas Palberg for
stimulating discussions. Moreover, we thank Norio Kikuchi for a critical
reading of the manuscript and Hans Knoth for his help in the preparation
of Figs.~\ref{fig9} and \ref{fig11}. We acknowledge financial support by the Deutsche
Forschungsgemeinschaft (DFG) under Grants No.~HO 2231/1 and HO 2231/2,
by the SFB 625 ``Von einzelnen Molek\"ulen zu nanoskopisch strukturierten
Materialien'', and by the SFB TR6 ``Colloidal Dispersions in External
Fields''.

\newpage
\section{List of Captions}

\fig{fig1}:
Sketch of the model for a colloidal particle. The centres of the small
spheres represent the points at which the colloidal particle interacts
with the LB fluid.  The cylinders that connect the spheres with each
other are just guides to the eye.

\fig{fig2}:
$1/\xi_{L}$ for particle of radius $R=2.5a$ as a function of $1/L$
for the indicated values of $\xi_0$. The dashed lines are fits with
Eq.~(\ref{eq_fitxi}) and the solid lines is a fit with Eq.~(\ref{eq_hasi})
for $\xi_0=19.8 m_0/\tau$ using the hydrodynamic radius $R_{\rm h}$
as a fit parameter.  The inset shows the variation of the hydrodynamic
radius $R_{\rm h}$ normalized by the assigned sphere radius $R$ as a
function of $\xi_0$ (see text).

\fig{fig3}:
$\xi_L$ as a function of the particle radius $R$ for $\xi_0=13.2 m_0/\tau$
and the system sizes $L=60a$ and $L=80a$.

\fig{fig4}:
Time--dependent velocity correlation function $C_v(t)$ (solid lines) of
the colloidal particle and normalized average velocity $v_{\rm s}(t)/V(0)$
of the LB fluid at the surface of the particle (dashed lines) for a)
$\xi_0=0.66 \; m_0/\tau$, b) $\xi_0=3.3 \; m_0/\tau$, c) $\xi_0=6.6 \;
m_0/\tau$, and d) $\xi_0=13.2 \; m_0/\tau$. The box length was set
to $L=40a$, the radius of the particle to $R=2.5a$, and its mass to
$M=120 m_0$. The dotted lines are exponential functions $f(t)=\exp\left(-
\frac{\xi_0}{M} t \right)$ that describe the short--time decay of $C_v(t)$
(note that no fit parameter is involved).

\fig{fig5}:
$C_v(t)$ for $\xi_0=6.6 \; m_0/\tau$ at the indicated system sizes. The
dotted line shows the power law $f(t) = A t^{-3/2}$ (see text).

\fig{fig6}:
The same as in Fig.~\ref{fig5} but now for the angular velocity
correlation function $C_{\omega}(t)$ and for $\xi_0=13.2 m_0/\tau$. The
dotted line shows the power law $f(t) = B t^{-5/2}$ (see text).

\fig{fig7}:
Velocity autocorrelation function for $\xi_0=6.6 m_0/\tau$ and $L=40a$
as calculated for a kicked particle (solid line) and with thermal
fluctuations (circles).

\fig{fig8}:
The same as in Fig.~\ref{fig7} but now for the angular velocity correlation
function.

\fig{fig9}:
Representative configuration of counterion and coion distribution around
a macroion of charge $Z_{\rm m}=255$.  The macroion is the central large
grey sphere, the $555$ counterions are the light gray coloured small
spheres and the $300$ coions are the black small spheres.

\fig{fig10}:
Plot of the radial density distribution $\rho_{\rm m}(r)$ of counterions
and coions around the macroion for the two indicated charges $Z_{\rm m}$.
The system with $Z_{\rm m}=255$ contains $555$ counterions and $300$
coions whereas there are $471$ counterions and $350$ coions in the system
with $Z_{\rm m}=121$.  The inset shows the fluctuation of the temperature
$T$ of the counterions around its assigned value (solid line).

\fig{fig11}:
Snapshot of the ionic distribution around the macroion (for the
system with $Z_{\rm m}=255$) with an applied electric field of
$E_x = 0.01$~V/\AA~(top) and $E_x=0.02$~V/\AA~(bottom), respectively.
The electric field is applied in the direction of the black arrow.
The rest is similar to Fig.~\ref{fig9}.

\fig{fig12}:
$N_{\rm c}^*/Z_{\rm m}$ as a function of the external electric field
$E_x$. $N_{\rm c}^*$ is the average number of counterions that move
with the particle. The inset shows the drift velocity $v_{\rm d}$ as a
function of the electric field $E_x$.

\newpage
\section{List of Figures}

\newpage

\begin{figure}[tb]
\includegraphics[width=0.75\columnwidth]{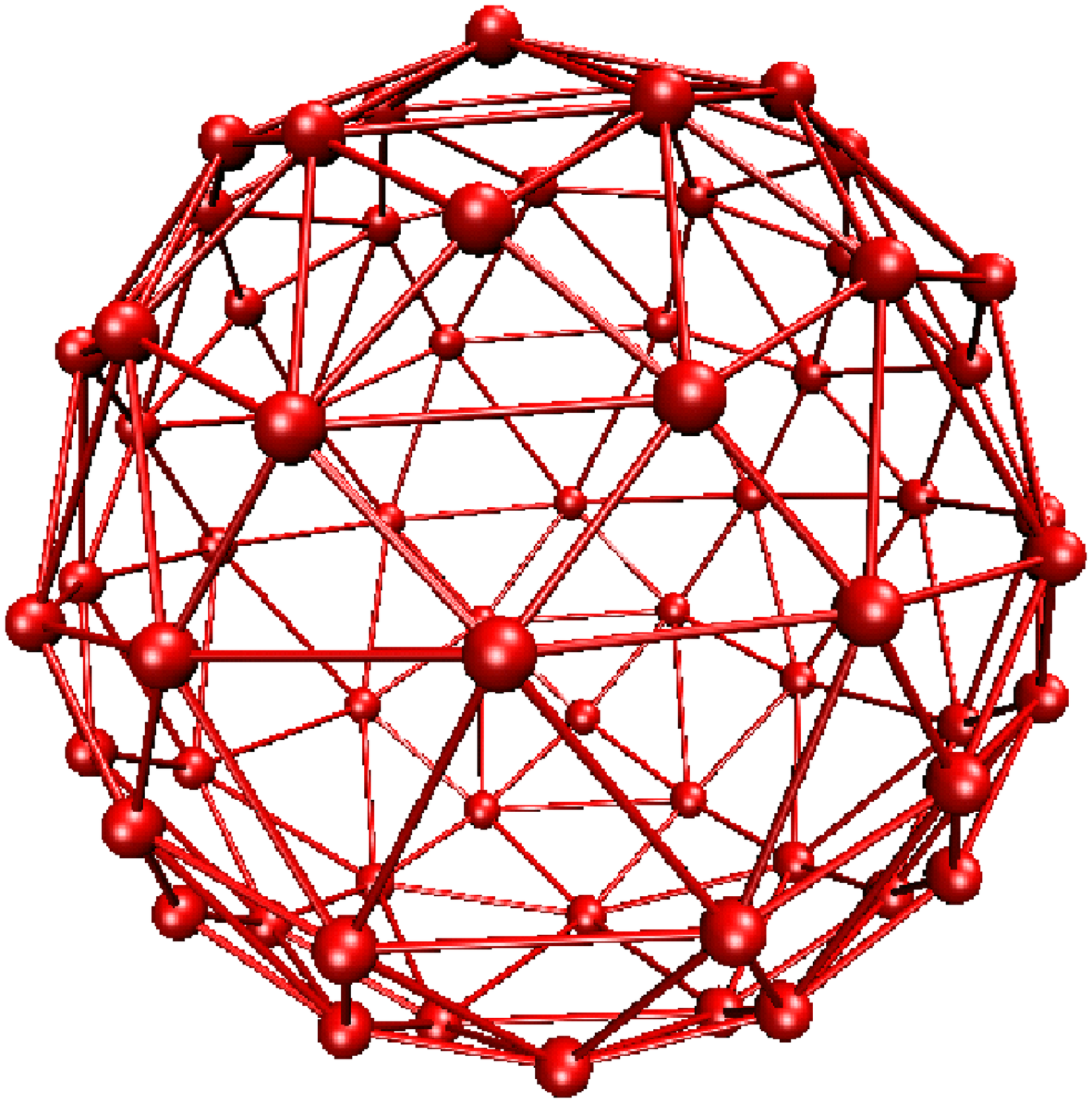}
\caption{\label{fig1}}
\end{figure}
\begin{figure}[tb]
\includegraphics[width=0.8\columnwidth]{fig2.eps}
\caption{\label{fig2}}
\end{figure}

\begin{figure}[tb]
\includegraphics[width=0.7\columnwidth]{fig3.eps}
\caption{\label{fig3}}
\end{figure}
\begin{figure}[tb]
\includegraphics[width=0.95\columnwidth]{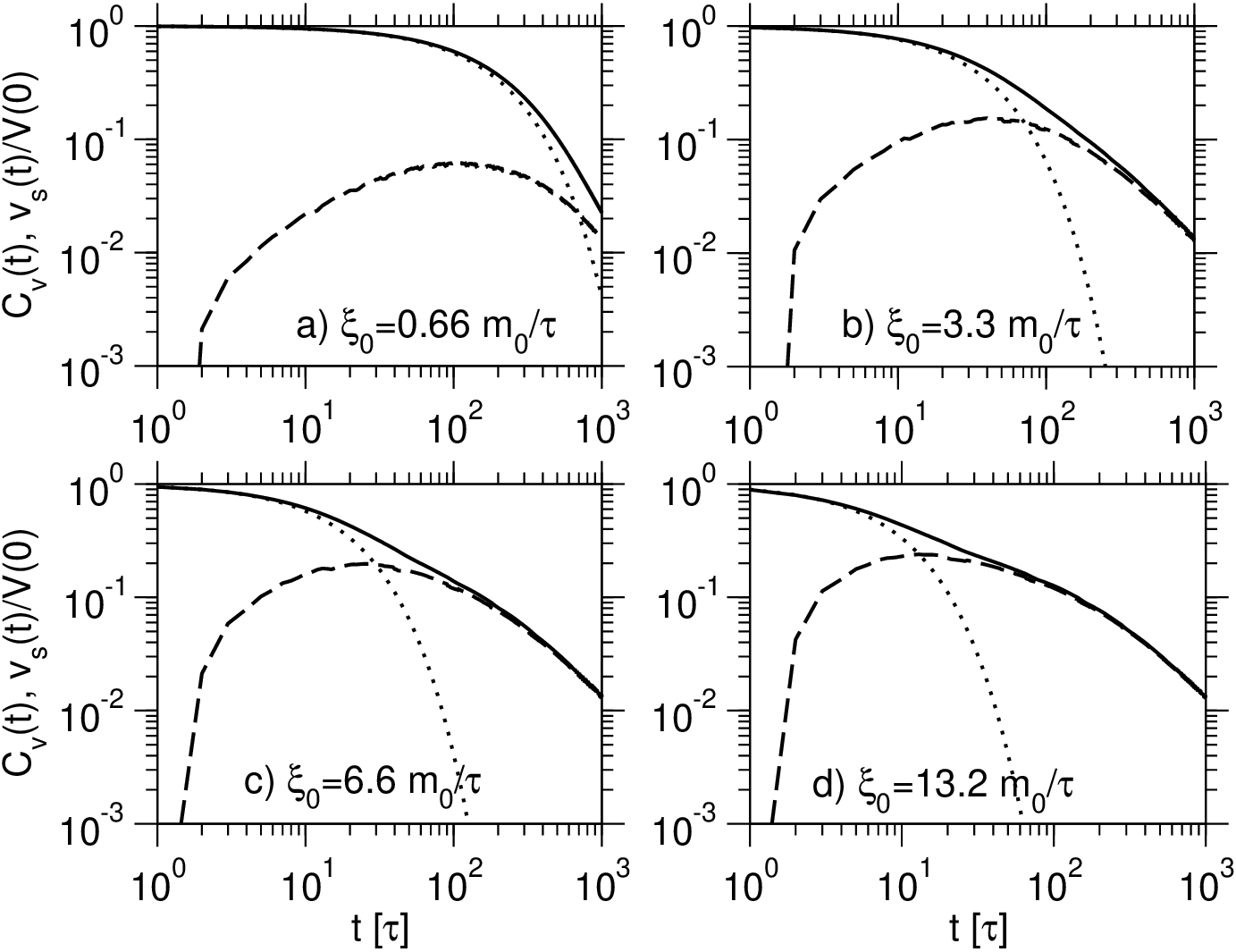}
\caption{\label{fig4}}
\end{figure}
\begin{figure}[tb]
\includegraphics[width=0.7\columnwidth]{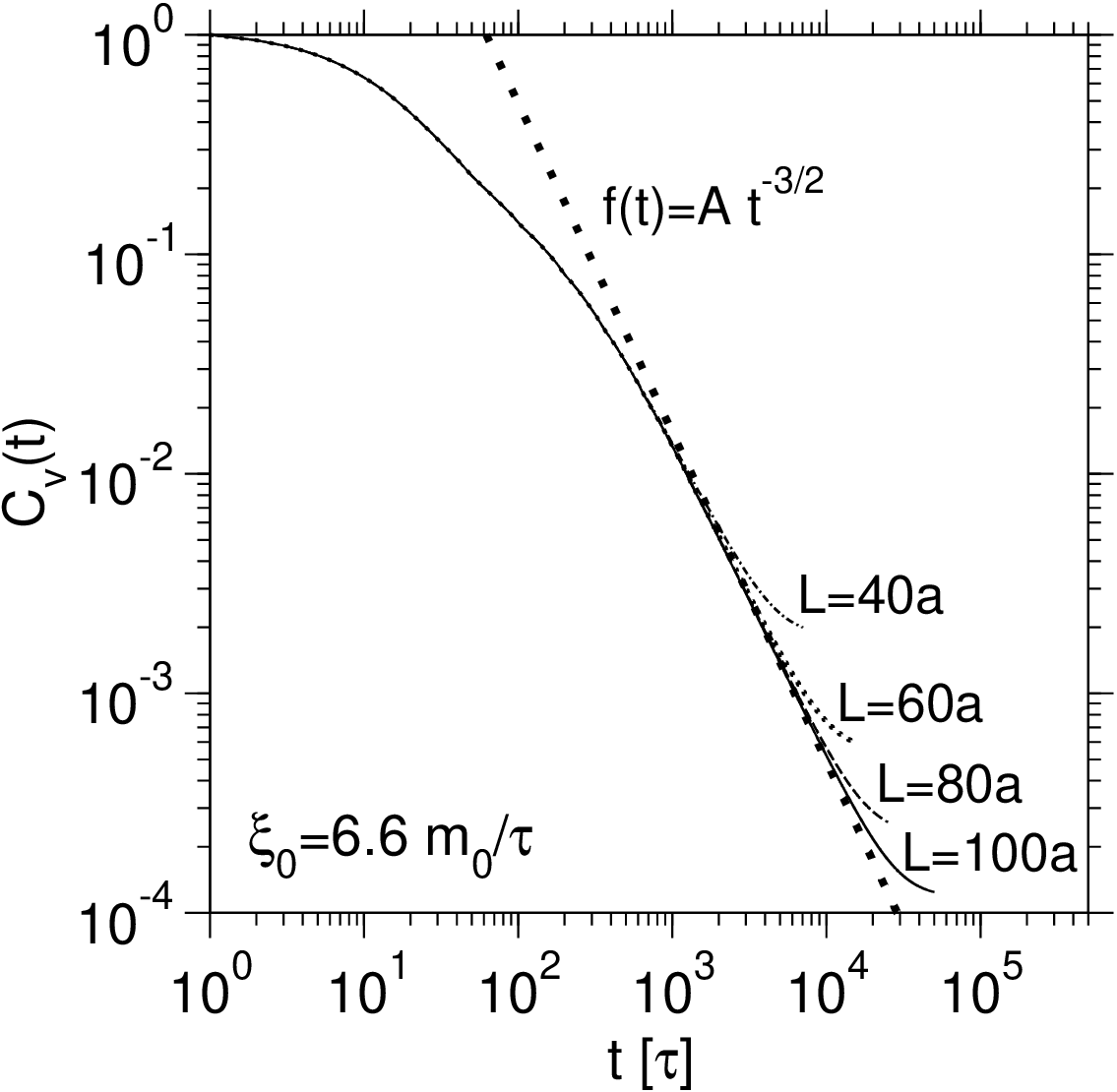}
\caption{\label{fig5}}
\end{figure}
\begin{figure}[tb]
\includegraphics[width=0.7\columnwidth]{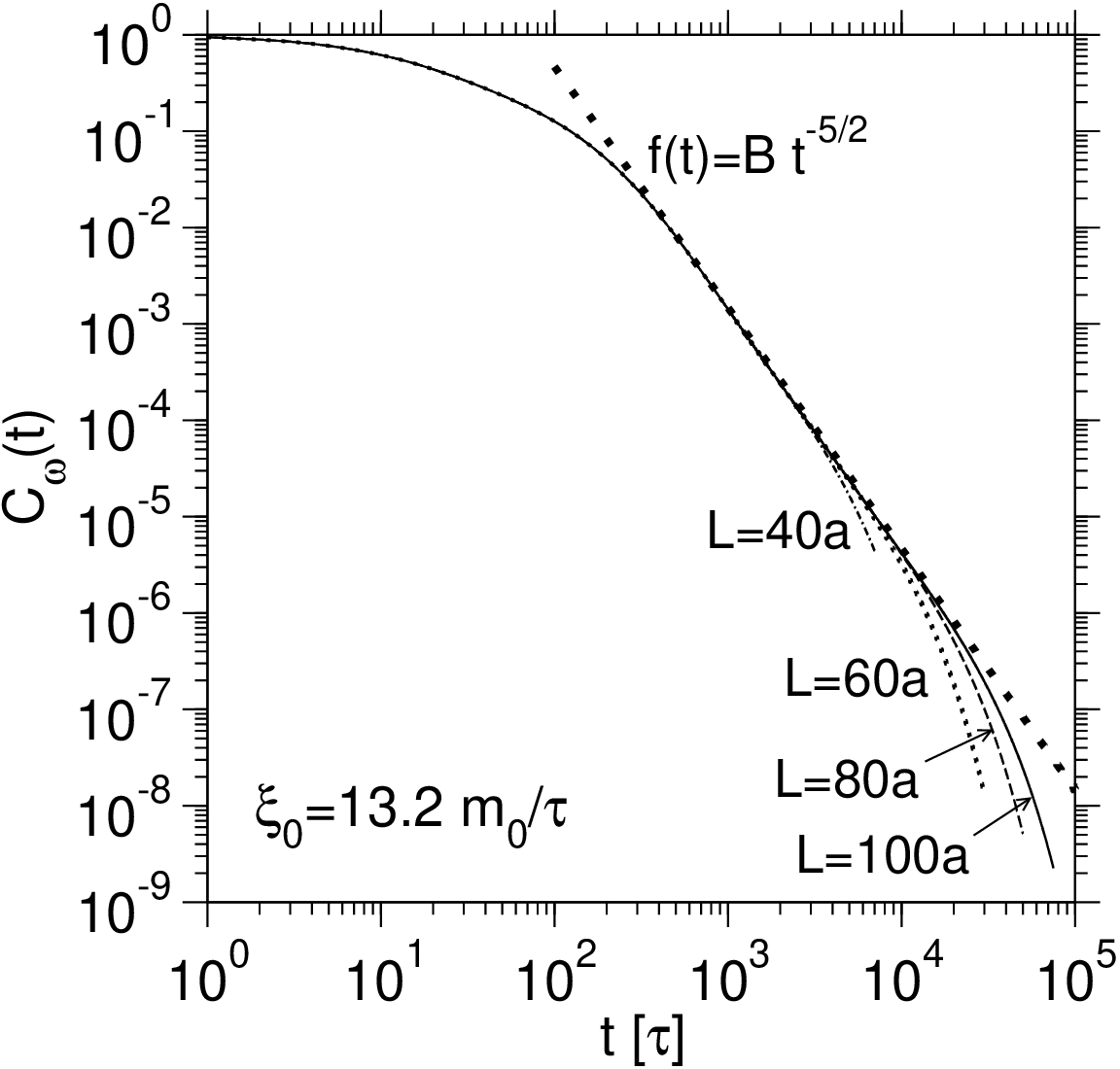}
\caption{\label{fig6}}
\end{figure}
\begin{figure}[tb]
\includegraphics[width=0.7\columnwidth]{fig7.eps}
\caption{\label{fig7}}
\end{figure}
\begin{figure}[tb]
\includegraphics[width=0.7\columnwidth]{fig8.eps}
\caption{\label{fig8}}
\end{figure}
\begin{figure}[tb]
\includegraphics[width=0.9\columnwidth]{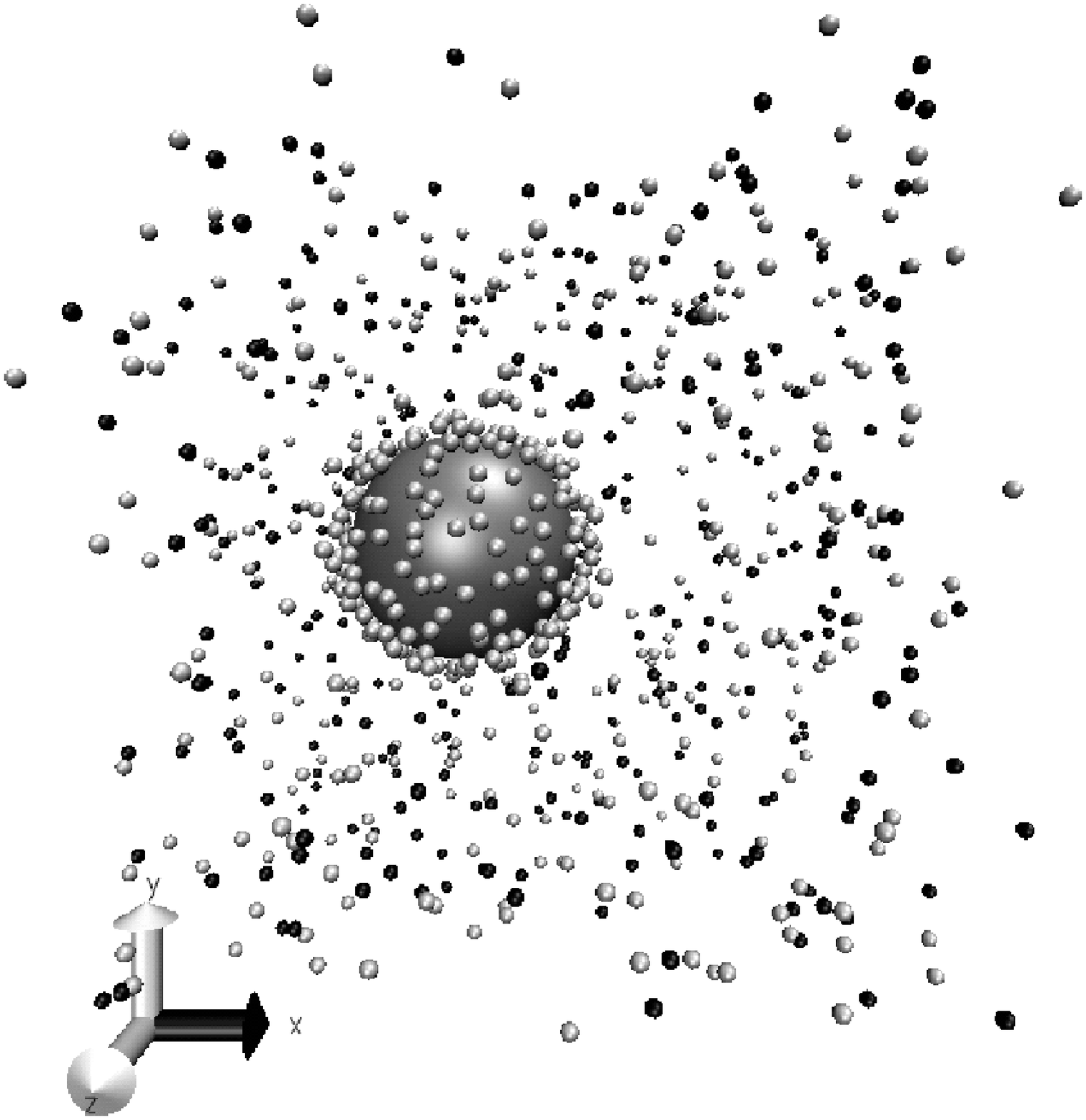}
\caption{\label{fig9}}
\end{figure}
\begin{figure}[tb]
\includegraphics[width=0.7\columnwidth]{fig10.eps}
\caption{\label{fig10}}
\end{figure}
\begin{figure}[t]
\includegraphics[width=0.6\columnwidth]{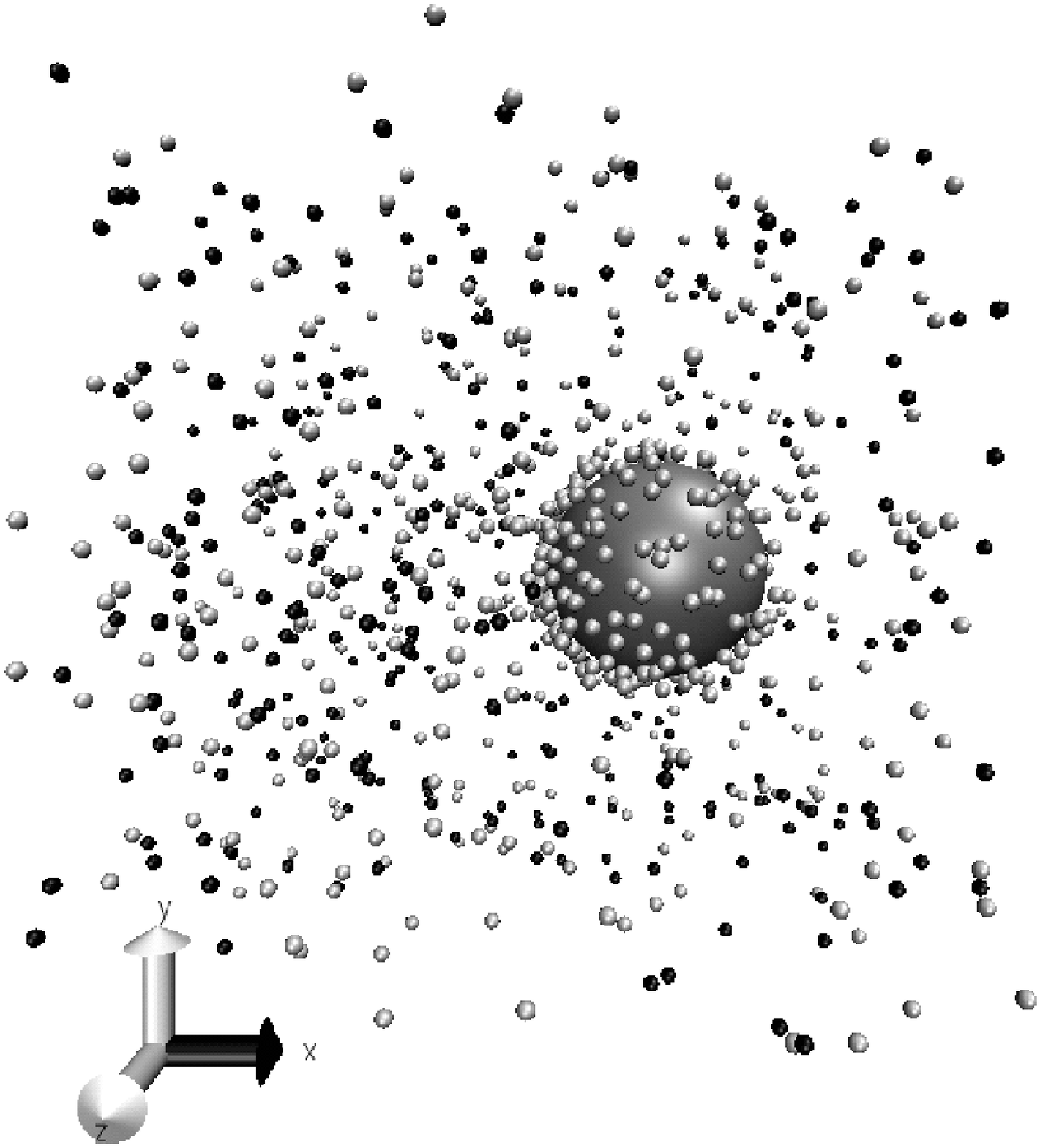}
\includegraphics[width=0.6\columnwidth]{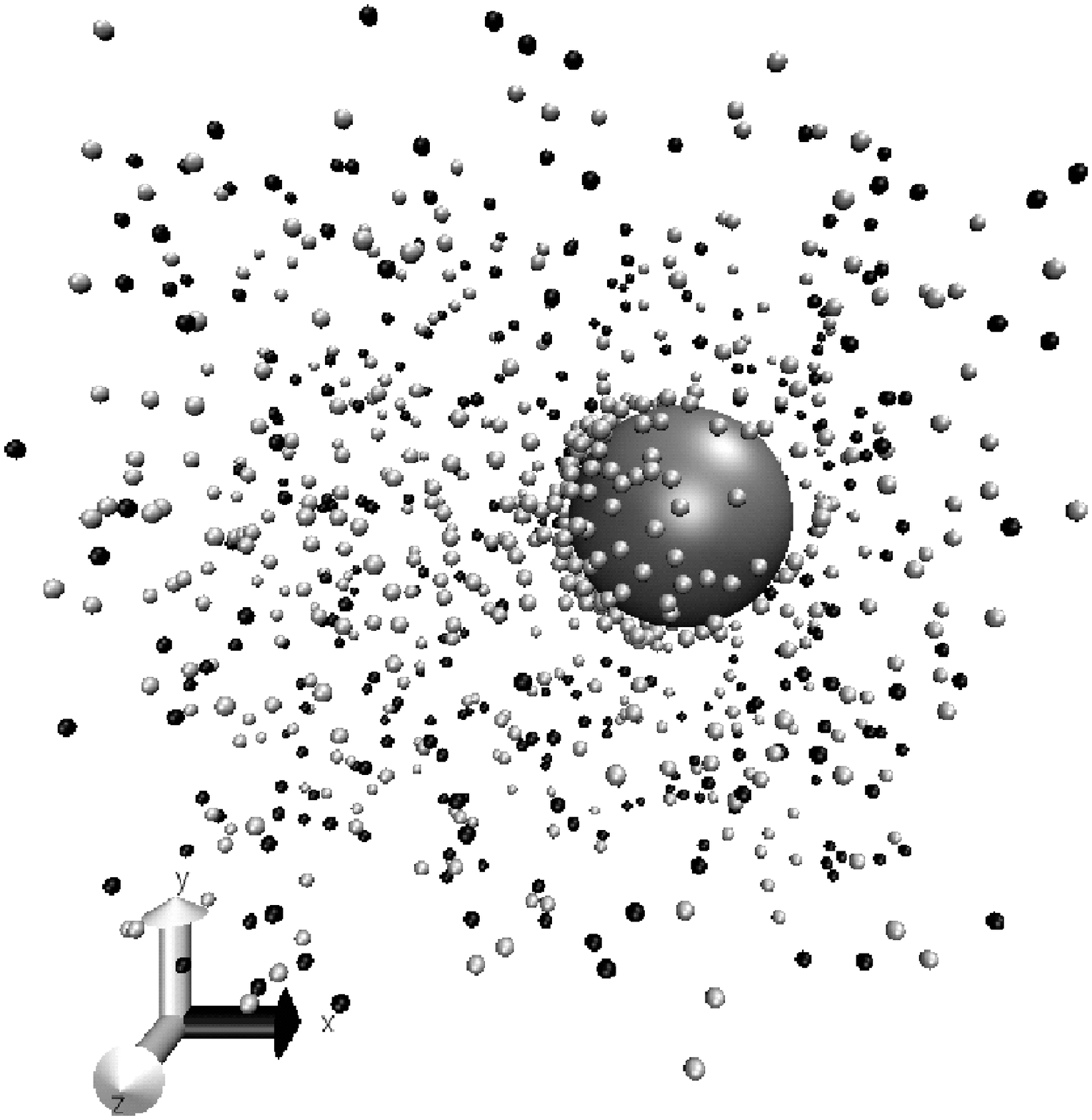}
\caption{\label{fig11}}
\end{figure}
\begin{figure}[tb]
{\includegraphics[width=0.7\columnwidth]{fig12.eps}}
\caption{\label{fig12}}
\end{figure}

\end{document}